\documentclass[12pt,preprint]{aastex}

\shorttitle{Variable Stars in the Globular Cluster NGC 6397}
\shortauthors{Kaluzny, \& Thompson}

\begin{document}

\title{ Time Series Photometry of Variable Stars  
in the Globular Cluster NGC 6397}

\author{J.~Kaluzny\altaffilmark{1}, and
I.~B.~Thompson\altaffilmark{2}}

\altaffiltext{1}{Copernicus Astronomical Center, Bartycka 18,
00-716 Warsaw, Poland; jka@camk.edu.pl}
\altaffiltext{2}{Carnegie Observatories, 813 Santa Barbara St.,
Pasadena, CA 91101-1292; ian@ociw.edu}

\begin{abstract}
Time series $BVI$ photometry is presented for 16 short-period variables
located in  the central region of the globular cluster NGC~6397. The
sample includes 9 newly detected variables. The light curve of
cataclysmic variable CV6  shows variability with a period of 0.2356
days.  We confirm an earlier reported period of 0.472 days for
cataclysmic variable CV1. Phased light curves of both CVs exhibit
sine-like light curves, with two minima occurring during each orbital
cycle. The secondary component of CV1 has a low average density of
0.83\ g\ cm$^{-3}$ indicating that it cannot be a normal main
sequence star.  Variables among the cluster blue stragglers include a
likely detached eclipsing binary with orbital period of 0.787 days,
three new SX~Phe stars (one of which has the extremely short period of
0.0215 days), and three low amplitude variables which are possible
$\gamma$ Doradus variables.

\end{abstract}

\keywords{ binaries: close - 
globular clusters: individual (NGC~6397) - novae - cataclysmic variables -
blue stragglers - stars:variables:other }

\section{INTRODUCTION}
The inner region of the nearby post-core-collapse globular cluster
NGC~6397 has been the target  of several surveys aimed at the
identification of rare or unusual stars likely to be created as the
result of stellar interactions in its dense cluster core (Coll \&
Bolton (2002) and references therein).  Recently Grindlay et al. (2001)
reported the detection with $Chandra$ of 25 X-ray sources within
2$\arcmin$ of the cluster center. Optical searches for variable stars
in the inner regions of globular clusters, particularly those with
strong central density cusps, have been hindered by crowding. Image
subtraction techniques in such strongly crowded fields have been
successful in locating and studying variables (Olech et al. 1999;
Kaluzny, Olech \& Stanek 2001), especially when the data are taken with
a fine plate scale with a stable and uniform point-spread function.
   
In this paper we present the analysis of ground-based time series CCD
photometry which was obtained to study the light curves of the optical
counterpart to the binary millisecond  pulsar PSR J1740-5340 (D'Amico
et al. 2001b; Ferraro et al. 2001). Our results for the binary
pulsar are given elsewhere (Kaluzny et al. 2002).  In this contribution
we report on  the results obtained for  other variable stars located in
the central part of the cluster.  Light curves for a total of 16
variables are presented and discussed.  Nine of these objects are new
identifications.
 
\section{OBSERVATIONS}

The photometric data were obtained with the 2.5-m du Pont telescope at
Las Campanas Observatory.  A field of $8.65\times2.60$~arcmin$^{2}$ was
observed with the TEK\#5 CCD camera at a scale of
0.259$\arcsec$/pixel.  This present  analysis is limited to a
sub-field $2.60\times2.60$~arcmin$^{2}$ centered approximately on the
cluster core. Most of the  data were obtained on 6 nights during the
period from UT May 1 to 8, 2002, with additional data obtained on UT
June 3, 2002.  Conditions were non-photometric on all but one night
with average seeing of 1.0$\arcsec$  in the $V$-band.

The cluster was observed for a total of 32 hours through $BVI_{\rm C}$
filters.  Exposure times were typically 30 sec ($B$), 15 sec ($V$), and
10 sec ($I$). Frames were were co-added, and the total number of
stacked images used in the present study was 69 ($B$), 196 ($V$), and
59 ($I$).  For a few of the variables discussed below we have also
extracted $V$-band time-series photometry from images with the best
seeing, selecting 1176  out of the total of 1256 available.  Our
observing material and photometric calibration procedure are described
in  detail in Kaluzny et al. (2002).

\section{IDENTIFICATION OF VARIABLES}
Two methods were used to detect potential variable stars.  Both of them
make use of the ISIS-1.2 image subtraction package (Alard \& Lupton
1998; Alard 2000).
 
The first method is based on examination of images created by combining
individual residual images, and relies entirely on tools included in
the ISIS package. This method is well suited for the detection of
variables with a high duty cycle \footnote{By "duty cycle" we mean the
fraction of time in which a given star shows luminosity other than its
median luminosity. An example of an object with low duty cycle is an
eclipsing binary with narrow eclipses.} and/or showing significant
changes of flux (bright variables or faint variables with large
amplitude light curves). An advantage of this method is that it permits
the detection of variables which cannot be resolved with classical
profile photometry in crowded fields.

The second method relies on the examination of the light curves of all
objects which could be measured on a reference frame with profile
fitting software. For each filter a reference frame was constructed by
averaging several stacked images of the best quality. The detection of
stellar objects and the extraction of  photometry was accomplished
using the DAOPHOT/ALLSTAR software package (Stetson 1987).  The total
of 4336 stars were measured on the $V$-band reference image. The
limiting magnitude depends very much on the distance from the cluster
core. The faintest measured stars have $V\approx 21.3$ and the observed
luminosity function starts to diminish at $V\approx 19.0$.  The $phot$
procedure in the ISIS package was used to extract differential light
curves at the position corresponding to each star detected with the
DAOPHOT/ALLSTAR package. Differential light curves were then
transformed to magnitudes and checked for variability.  The light
curves were searched for the presence of any periodic signal with the
$AoV$ algorithm (Schwarzenberg-Czerny 1989; Schwarzenberg-Czerny 1996)
and were reviewed for possible eclipse-like events.
  
A total of 16 variables were identified. Nine of these are new
detections. Information on these variables is presented in Table 1.
Column 1 gives an assigned name following Clement et al. (2001),
followed by the right ascension and declination of the variable
\footnote{Clement et al. (2001) list 12 variables in NGC 6397, V1-V12.
Four variables listed in Table 1 are present in the Clement et al.
paper, and we copy their name assignments (V7, V10, V11 and V12).
Variables  V13-V24 are not in the Clement et al. catalogue.}. Column 5
gives alternate designations for the variables, many of which have been
previously identified as blue stragglers, UV bright objects, and/or
candidate cataclysmic variables.

Table 1 also gives positional information for all 16 variables. Columns
2 and 3 list equatorial coordinates, while column 4 lists pixel
coordinates of the variables as found on $HST$ archive image u5dr0401r
for 12 of the 16 variables. The equatorial coordinates were derived
from the frame solution included in the $HST$ image header.  We derived
equatorial coordinates for the four variables laying outside the $HST$
field from an astrometric solution to our $V$ band reference image
based on 1022 stars with equatorial coordinates adopted from Kaluzny
(1997).  This solution has residuals not exceeding 1$\arcsec$ when
compared to the $HST$ astrometric solution.  Finding charts for
variables V10, V22 and V23 can be found in Lauzeral et al. (1992; stars
11, 8 and 16 in their Fig. 1), and a finding chart for variable V24 is
shown in Figure 1.

Basic information on the photometric properties of the 16 detected
variables is presented in Table 2. Column 1 lists the variable name,
followed by a classification of the light curve, the period
of variability, mean $B$, $V$, and $I$ magnitudes, and the $V$-band
full amplitude of variability. Positions of the variables in the
cluster color-magnitude diagram are shown in Fig. 2.  The ellipsoidal
variable V16 is the optical counterpart to the millisecond binary
pulsar J1740-5340 (D'Admico et al. 2001; Ferraro et al. 2001).
Photometry of this variable is discussed in Kaluzny et al. (2002).

\section{ANALYSIS OF PHOTOMETRY}

\subsection{Cataclysmic Variables}

Grindlay et al. (2001)  detected  9 possible cataclysmic variables (CV) in
the central region of NGC~6397 with the $Chandra$ telescope, naming the
objects CV1 - CV9.  They identified the optical counterparts of CV1 -
CV5 based on observations reported by Cool et al. (1995, 1998), Grindlay (1999), and unpublished HST H-$alpha$ observations (Grindlay et al. 2001).


Variable V12 can be unambiguously identified with CV1 using positional
data provided in Cool et al. (1998; see their Table 1).  From an
examination of the HST archive image  u5dr0401r at the equatorial
coordinates of V13 we conclude that this star is the optical
counterpart of CV6. On the $HST$ image V13 is visible as an isolated
object with a closest neighbor at a distance of $d\approx
0.8\arcsec$.

Variables CV2 - CV5 and CV7 - CV9 could  not be resolved on our
reference images.  We attempted to extract  differential light curves
using the ISIS package at the known positions of these stars.  The
light curves suffered from large photometric errors and showed no sign
of any periodicity. The large errors  result from the effects of
several relatively bright stars near the variables.

\subsubsection{CV1}

Examination of the light curves of V12~=~CV1 reveals the presence of a
sine-like  periodic modulation.  The power spectrum calculated from the
$V$ filter time series photometry based on individual exposures is
presented in Fig. 3.  Two maxima  of comparable strength
are present at frequencies corresponding to periods of 0.4712  and
0.2356 days.  Grindlay et al. (2001) report that CV1 showed one total
eclipse in X-rays through the 0.567 days observation obtained with
$Chandra$, and therefore we may eliminate the shorter period from
consideration.  Nightly light curves of CV1 phased with the period of
0.4712 days are displayed in Fig. 4. The average value of the formal error
of a single data point is 0.042~mag.  It is worth noting that the
shape of the light curve as well as the average luminosity of the system
are relatively stable over the interval covered by our observations.
Figure 5 shows phased $BVI$ light curves corresponding to photometry
extracted from averaged images. Neither  the amplitude nor the shape of the
light curves change noticeably with band-pass.  The light curves
are quite symmetric with two maxima of comparable height separated by
half of the period. The two observed minima have comparable depths
although the minimum occurring at phase zero is slightly sharper. These
properties of the light curves of CV1 suggest that the observed variability
is dominated in optical domain by the ellipsoidality effect
caused by rotation of the Roche lobe filling component, and that the 
\footnote{For cataclysmic variables it is conventional to call the 
degenerate component the "primary" and its companion the "secondary", independent of the actual mass ratio of a given system.}
secondary dominates optical flux of the system. Such an interpretation
is consistent with the relatively red colors of CV1. The variable is
located about 0.1 mag to the blue of the cluster main sequence on the
$V/B-V$ plane and it is located slightly to the red of the main
sequence on the $V/V-I$ plane (See Fig. 2).  At the time of our
observations the optical flux generated by the accretion process
apparently contributed a small fraction of the total optical luminosity
of the binary.

Another interesting property of CV1 is its relatively long orbital
period.  Among 318 cataclysmic variables  which are listed in  Ritter
\& Kolb (1998) there are only 14 objects with periods longer than 0.47
days.  It is possible to get a robust and reliable estimate of the
average density of the secondary component of CV1 from the formula in
Faulkner et al. (1972) and Eggleton (1983):
  \begin{eqnarray}
 <\rho > = 107P^{-2}
   \end{eqnarray}
where the period P is in hours and the density $\rho$ is in ${\rm
g\ cm^{-3} }$. We obtain $<\rho>=0.83$~${\rm g\ cm^{-3}}$. The binary is
located well below the cluster turn-off and therefore we may expect that
mass of the secondary does not exceed $0.8M_{\odot}$. Theoretical 
models published recently by
Bergbusch \& VandenBerg (2001)  give an average density
$<\rho>=3.52$~${\rm g\ cm^{-3}}$, for a ZAMS model of a 0.8 solar mass
star with ${\rm [Fe/H]=-2.0}$.  For a mass lower than 
$0.8$~m$_{\odot}$ the expected density is even higher  as
$<\rho>\sim m^{-2}$.  We conclude that the secondary
component of CV1 is noticeably over-sized compared to a normal low-mass
main-sequence star.

Knowing the distance modulus of the cluster we may derive an absolute
luminosity of the variable. For $(m-M)_{\rm V}= 12.69\pm 0.15 $ and
$E(B-V)=0.18$ (Reid \& Gizis 1998) we obtain $<M_{\rm V}>=4.7$ for the
average absolute magnitude of CV1 in the $V$-band. It is tempting to
use that information to derive the radius of the secondary star but we
feel that uncertainties in  relative intensity of the accretion
generated flux to the total observed luminosity are too large.  Such
uncertainties affect not only the estimate of the flux from  the
secondary but also any estimate of its effective temperature.  These
problems can be greatly reduced by observing the binary at near-IR
wavelengths where the total  flux of the system should  be strongly
dominated by the secondary star.

We conclude this part of the discussion by noting that  X-ray observations
presented by Grindlay et al. (2001) are consistent with the
identification of CV1 as either an ordinary dwarf nova or a magnetic CV.

\subsubsection{CV6}

The power spectrum of the $V$ band light curve of V13~=~CV6 shows
strong peaks at periods of 0.1176 and 0.2352 days.  An examination of
the light curves from individual nights indicates that they  exhibit
two minima of different shape separated by about 0.12 days.  This is
particularly clear in the light curve extracted from individual images
collected on  the night of UT May 1, 2002,  which is presented in Fig.
6. The $BVI$ light curves of CV6 phased with a period of 0.2352 days
are shown in Fig. 7. They are based on photometry extracted from
averaged images.  The shape and mean level of the light curves were
quite stable during our observations. As for CV1 the variability of CV6
seems to be dominated by the ellipsoidality effect.  The minimum
occurring at  phase 0.0 is narrower than the minimum at phase 0.5. We
interpret this as  evidence that the bright accretion region
surrounding the primary component of the binary is eclipsed at phase
0.0.  The variable is located about $0.15~mag$ to the blue of the
cluster main sequence in the  $V/B-V$ plane (see Fig. 2) and in the
$V/V-I$ plane it is located  at the red edge of the cluster main
sequence.

Assuming cluster membership for CV6 we estimate $M_{\rm V}=6.2$.
The average density of the secondary component is
$\sim$ ${\rm 3.4\ g\ cm^{-3}}$, consistent with the density expected 
for a slightly evolved Pop II main sequence star of mass  
$0.7\ m_{\odot}$. In particular, models published by 
Girardi et al. (2000) predict $<\rho>={\rm 3.2\ g\ cm^{-3}}$
and $M_{\rm V}=6.2$ for 0.7 solar mass star of age 11~Gyr.

\subsection{Eclipsing Binaries}
Our sample of variables includes four eclipsing binaries.
In this section we comment briefly on their properties,
a detailed analysis will require spectroscopic observations.

Variable V7 was identified as a W UMa variable by Kaluzny (1997).  This
star  has two close visual companions with $V=18.31 $ and $V=19.07$
located at angular distances of $0.54\arcsec$ and $0.56\arcsec$,
respectively. Despite the proximity of the companions the pixel scale 
of the observations together with the high signal-to-noise data
meant that photometry could be measured for both companions  in our 
$V$ band data.  Only the brighter companion could be measured
while extracting photometry for the $B$ and $I$ bands. Both companions
were unresolved in the photometry reported by Kaluzny (1997), leading
to an overestimation of the luminosity of V7 as reported in that paper.
Variable V19=PC-1 was identified by Taylor et al. (2001) in a
photometric survey for objects with  excess ${\rm H}\alpha$ flux. They
also detected V7 in the course of their survey. V19 has a close visual
companion of $V=17.22$ at angular distance of $0.5\arcsec$. That
companion is measured in our photometry for all bands. 

Phased $V$-band light curves of V7 and V19 are shown in Fig. 8.  Some
intrinsic night-to-night changes of the shape of the light curve were
observed for V7. Such behavior is not unusual for W~UMa type systems.
There is some indication that the secondary minimum of V7 exhibits a
"flat-bottom" indicating that this eclipse is total.  For W~UMa type
systems with total eclipses one may obtain reliable light curve
solutions as totality removes the degeneracy between the mass ratio and
inclination of the system (Mochnacki and Doughty 1972).  These two
contact binaries  have similar colors and lie about $0.7$~mag above
cluster main sequence on the color-magnitude diagram, suggesting that
that they are most likely members of NGC~6397.

Variable V14 is a relatively faint eclipsing binary located about
0.1~mag to the red of the cluster main sequence.  Its phased light
curve is presented in Fig. 9,  two shallow minima of different depth
are seen.  Main sequence binaries with periods below 0.35 days almost
always show ordinary EW type light curves, a signature of a contact
configuration.  If V14 is a member of NGC6397 then its red color and
faint magnitude would suggest the components are of late spectral type
with low masses and radii. In this case V14 could be a close  but 
non-contact binary despite its short period.

Variable V18 is by far the most interesting of the four eclipsing
binaries included in our sample. It is not only a likely blue straggler
but it also shows a very unusual light curve (see Fig. 10). At first
glance it resembles the light curves of ordinary W UMa type contact
binaries. However the light curve of V18 shows clear signatures of
eclipse ingress and egress, not observed in  contact binaries. We
conclude that V18 is a detached or semi-detached system composed of
stars with very similar surface brightness.  The light curve is similar
in all 3 filters with some indication that eclipses become progressively
shallower from the $B$ to the $I$ band by $\sim0.02 $ mag.

Examination of $HST$ images of the cluster shows  that V18 possesses a
close visual companion at an angular distance of 0.19~$\arcsec$.  It
has not been resolved in our profile photometry and hence its flux
acts as a "third" light in the photometry of V18.  We have identified
the companion as star 200338 in the data base published by Piotto et
al. (2002), with $V=18.631$ and $B=19.338$. We adjust these magnitudes
to $V=18.613$ and $B=19.281$ to take into account differences in the
zero points of the two sets of photometry of $-0.018$ and $-0.057$ for
the $V$ and $B$ filters, respectively (our magnitudes are brighter).
Since $I$-band data are not included in the Piotto et al. (2002) study
we estimate $I$-band photometry from the  fact that the V18 companion
lies on the cluster main sequence.  From  the $VI$ photometry of
NGC 6397 published by Alcaino et al. (1997) we estimate that for
$V=18.61$ the companion has $I\approx 17.52$.  Light curves presented
in Fig. 10 as well as magnitudes listed for V18 in Table 2 are
corrected for the contribution of the nearby companion.

Attempts to derive a reliable light curve solution for V18 are hampered
by the lack of vital information on the mass ratio of the binary.  We
have calculated a grid of solutions for a wide  range of assumed values
of the mass ratio $q=m_{2}/m_{1}$. Index "1" refers to star eclipsed at
phase "0".  The $V$-band light curve was was solved using
Wilson-Devinney code (Wilson 1979) embedded in the MINGA minimizing
package (Plewa 1988).  Our preliminary results can be summarized as
follows.  For $0.08<q<2.35$ solutions imply a detached configuration
with an inclination in the range $61<i<74$~deg.  The ratio of component
radii is constrained to the range $0.83<r_{2}/r_{1}<1.17$.  For
$q>2.35$ solutions converge to semi-detached configurations with the
less massive component filling its Roche lobe. The $\chi^{2}$ statistic
measuring the quality of fit of the synthetic light curve to the
observations has a minimum near $q=0.20$. A solution for that specific
value of the mass ratio gives inclination $i=71$~deg and average
relative radii of the components $r_{1}=0.22$ and $r_{2}=0.24$.  If the
mass ratio is  close to 0.2 and the system is detached, then one may
wonder why both components have very similar effective temperatures.
Note that the  color of the variable is essentially constant over the
whole orbital period.  However, if the mass ratio is close to unity then
we face difficulty trying to explain how this blue straggler can be
composed of two stars both of which are significantly bluer and more
luminous than stars at the cluster turnoff. Spectroscopic data
providing information about mass ratio of the binary are needed to
reliably determine its geometrical and absolute parameters.
 

\subsection{SX~Phe Stars}
SX~Phe type variables are short period pulsating stars which can be
considered Pop~II counterparts of more metal rich $\delta$~Sct type
stars. It is not unusual to find them among blue stragglers in globular
clusters (Rodriquez \& L\'opez-Gonz\'ales 2000).  Variables  V10 and
V11 were originally identified as SX~Phe stars by Kaluzny (1997).  Here
we add three more objects to that group.  Light curves of all five
variables show modulation of shape and amplitude indicating the
presence of multimodal pulsations.  A detailed analysis of these data
will be published in a separate paper  (Schwarzenberg-Czerny et al.; in
preparation).  Here we note only that variables V10 and V15 have
extremely short dominant periods, at 0.0215 days V15 has the shortest
period known for an SX~Phe star.  No SX~Phe stars with periods below
0.030 days are  listed in the recently published catalog of Rodriquez
\& L\'opez-Gonz\'ales (2000).  We considered the possibility that V15
is  not an SX~Phe type star but rather a pulsating hot subdwarf.
However its $B-V$ color is too red to be a  bright sdB/O star (note the
position of V15 in Fig.  2).

\subsection{Other Variables}

In this section we discuss briefly four variables which 
cannot be classified with confidence based on the available data.

Variable V20 is  one of the brightest blue stragglers identified in the
cluster core by Lauzeral et al. (1993).  The power spectrum of its
light curve shows two major peaks at periods of 0.861 and 0.436 days,
with the higher peak corresponding to the longer period. The $V$ band
light curves of V20 phased with each of these two periods are shown in
Fig. 11.  For the longer period the light curve has two minima, and
this suggests that V20 is a low amplitude W UMa type system. The
feature visible at the second quadrature arises from a different light
level observed on the single night of June 3, roughly one month after
the first observing run when most of the data were collected.  The
period of 0.861 days is relatively long for a contact binary belonging
to a globular cluster (Rucinski 2000).  Yet another possibility is that
V20 is a close binary with variability due to the ellipsoidality
effect.  The light curve phased on the shorter period is slightly more
noisy with a single minimum. The period of 0.436 days is far too long
to classify V20 as a pulsating SX~Phe star.  The variable is too hot to
show spot-related activity as is observed for FK~Com or BY~Dra type
stars. However, it can  be related to $\gamma $~Doradus stars, as it is
discussed below for two other variables.

Phased $V$-band light curves for variables V17 and V24 are presented in
Fig. 12. They both show low amplitude, sine-like  modulation, with
periods of  0.457 days and 0.525 days for V24 and V17, respectively. On
the color-magnitude diagram the stars are located about 0.15 mag to the
red of the cluster turnoff.  We propose that V17 and V24 are Pop II
counterparts of $\gamma$~Doradus variables. The $\gamma$~Doradus stars
have often  multiple periods between 0.4 and 3 days and show sinusoidal
light curves with amplitudes -- in optical domain -- of the order of
0.01 mag (Zerbi 2000; Henry \& Fekel 2002). Their variability is due to
non-radial $g$-mode pulsations and they are usually subgiants or, less
frequently, main sequence stars of spectral type F0-F2.  The red edge
of the instability strip for Pop I $\gamma$~Doradus variables is
located at $(B-V)\approx 0.37$ (Henry \& Fekel 2002). The dereddened
colors of V17 and V24 are $(B-V)_{0}=0.45$ and $(B-V)_{0}=0.37$,
respectively.  Henry \& Fekel (2002) note that $\gamma$~Doradus
variables show average $A(B)/A(V)$ amplitude ratios of $\sim$~1.3. This
distinguishes them from "spotted" variables which have  amplitude
ratios of $\sim$~1.1 and from ellipsoidal variables for which the ratio
is $\sim$~1.0.  From our data we obtain $A(B)/A(V)=1.09\pm 0.21$ and
$A(B)/A(V)=1.25\pm 0.14$ for V17 and  V24, respectively.  More extended
time series would allow a more accurate estimate of  the $A(B)/A(V)$
ratio for both variables and would also help to search for
multiperiodicity in the light curves. The stability of $\gamma$~Doradus
light curves 100-200 cycles also  distinguishes these stars  from
"spotted" variables.

Variable V22=BS8 is a bright blue straggler which was identified in the
cluster  core by Lauzeral et al. (1993). The light curve of V22 cannot
be phased with a single period although during May run  we observed
four minima occurring in 1 and 2 day intervals. The light curve
extracted from observations obtained on nights of 9 and 10 July, 1995
(Kaluzny 1997) shows variations with $\delta V\approx 0.12$ and a
possible period of about 0.75 days. However that period does not fit
the 2002 data. It is possible that V22 is a pulsating multiperiodic
variable related to $\gamma$~Doradus stars or that it is a distant
RR~Lyr variable of RRd type. In Fig. 13 we show light curves from two
nights on which the observed variations were most pronounced.

\section{SUMMARY AND CONCLUDING REMARKS}

We have used time series photometry obtained with a medium sized
telescope to look for short period variables in the central part of
the post-core-collapse cluster NGC~6397. We show that by applying the
image subtraction technique that it is possible not only to detect
variable stars in very crowded fields but also to measure accurate
light curves for objects with amplitudes as small as 0.01 mag.
Photometry of NGC 6397 obtained with $HST$ imaging  (Piotto et al.
2002) allows a check and, if necessary, a correction  for contamination
from possible visual companions which are unresolved in ground-based
data.

We present the first complete light curves and derive orbital periods
for two cataclysmic variables in NGC~6397. The total flux
and variability of both of these CV's is dominated by the secondary
components.

Several new variables have been  identified among the cluster blue
stragglers, including one detached binary and three objects  being 
possible Pop II counterparts of $\gamma$~Doradus stars.

The determination of cluster membership of the detected variables has
relied on  their positions
in the cluster CMD. While the
cluster is located on the sky near the Galactic bulge at $l=338$~deg
and $b=-12$~deg, in the central region cluster stars must prevail
strongly. However we cannot exclude the possibility that some of the
variables are field interlopers. The publication of proper motion
catalogs (Cool \& Bolton 2002) will be exceptionally useful in
resolving issues of cluster membership.

\acknowledgments

JK was supported by the Polish KBN grant 5P03D004.21 and by
NSF grant AST-9819787. IT was supported by NSF grant AST-9819786.
We would like to thank Alex Schwarzenberg-Czerny for 
providing us with his excellent period finding programs.

\clearpage

\begin{figure}
\figurenum{1}
\epsscale{.4}
\plotone{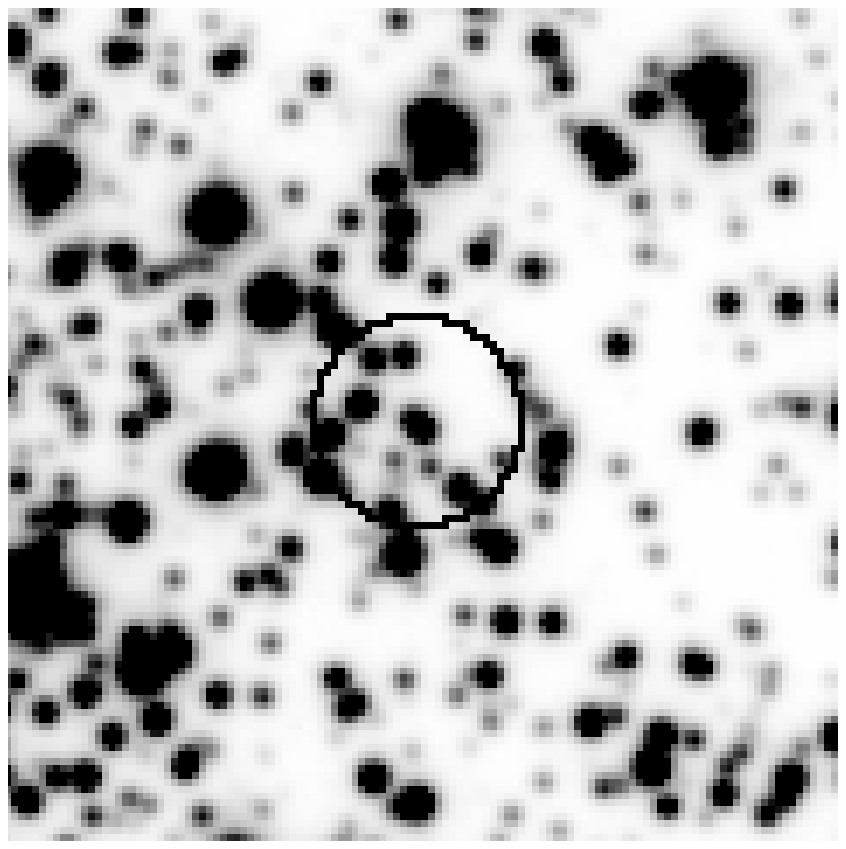}
\caption{A finding chart for variable V24. North is up
and east is to the left. The field of view is 30~$\arcsec$
on a side. The variable is the W-S component of a blend.}
\end{figure}

\clearpage

\begin{figure}
\figurenum{2}
\epsscale{1.}
\plotone{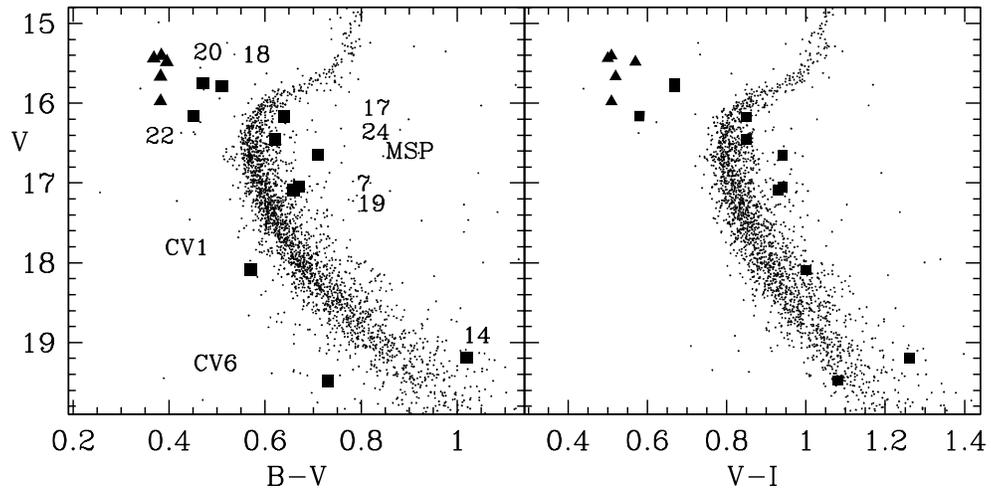}
\caption{Color-magnitude diagrams of NGC~6397 with the positions of the
variables marked. SX~Phe stars are labeled with triangles. Eclipsing
binaries and CVs are labeled on the left figure. For SX~Phe stars and
CVs the positions correspond to average colors and magnitudes.
}
\end{figure}

\clearpage
\begin{figure}
\epsscale{1.0}
\figurenum{3}
\plotone{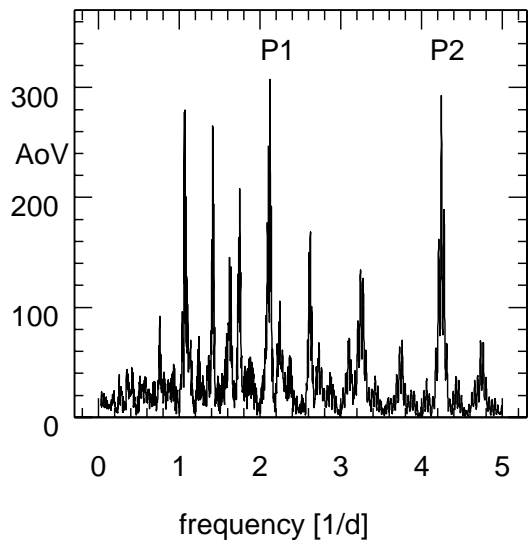}
\caption{The power spectrum for the $V$-band light curve of the variable V12=CV1.}
\end{figure}

\clearpage

\begin{figure}
\figurenum{4}
\plotone{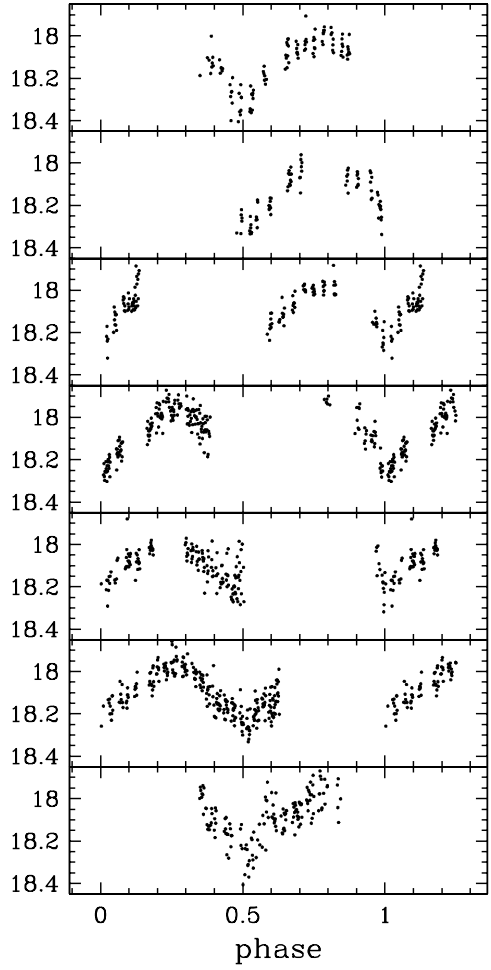}
\caption{Phased $V$-band light curves of variable V12=CV1 for
nights of UT May 1, 2, 3, 5, 6, and 7, and UT June 3, 2002  (from top to   
bottom).
}
\end{figure}

\clearpage

\begin{figure}
\figurenum{5}
\plotone{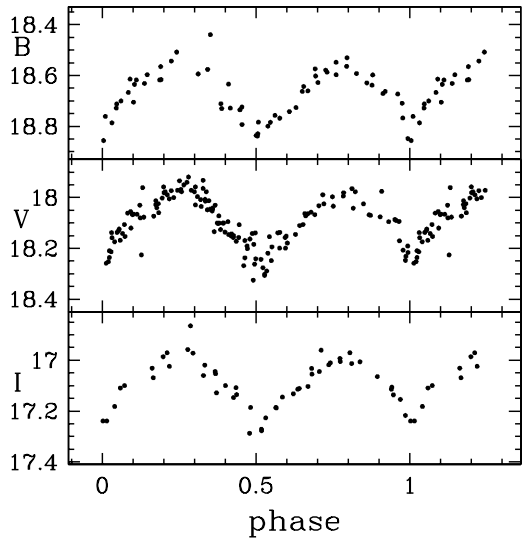}
\caption{Phased $BVI$ light curves of variable V12=CV1. Note 
the same scale for all filters.}
\end{figure}

\clearpage

\begin{figure}
\figurenum{6}
\plotone{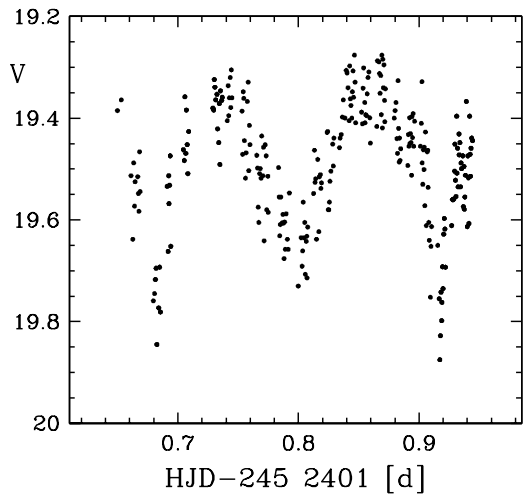}
\caption{Time domain $V$-band light curve of V13=CV6  for the night of UT May 1, 2002}
\end{figure}

\clearpage

\begin{figure}
\figurenum{7}
\plotone{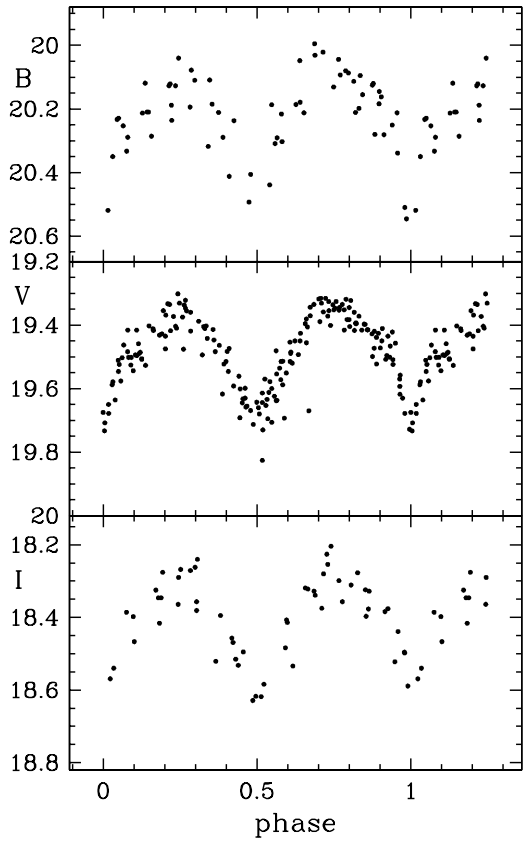}
\caption{Phased $BVI$ light curves of variable V13=CV6}
\end{figure}

\clearpage
\begin{figure}
\figurenum{8}
\plotone{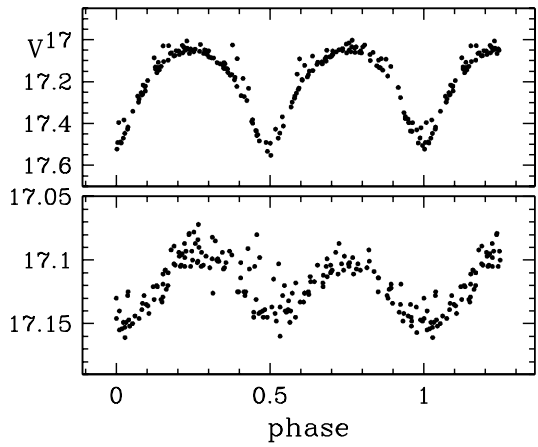}
\caption{Phased $V$-band light curves of variables V7 (upper)
and V19.}
\end{figure}

\clearpage
\begin{figure}
\figurenum{9}
\plotone{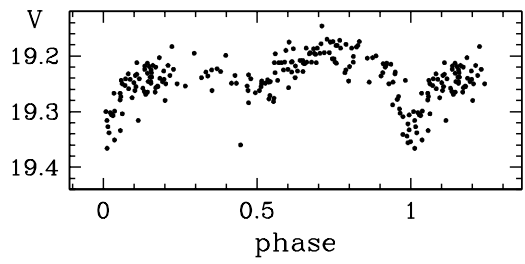}
\caption{Phased $V$-band light curve of variable V14}
\end{figure}

\clearpage

\begin{figure}
\figurenum{10}
\plotone{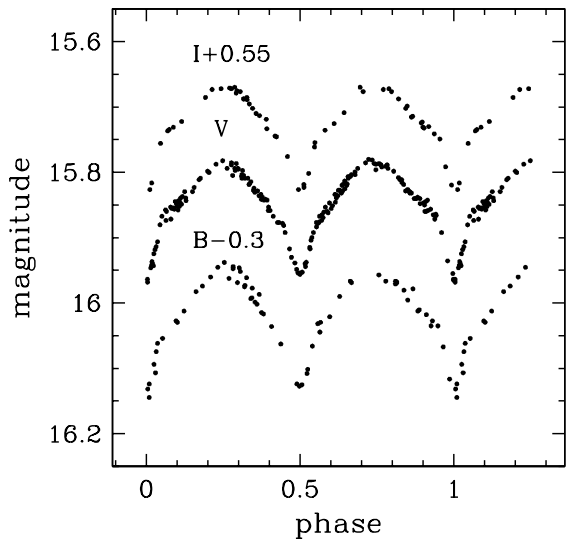}
\caption{Phased $BVI$ light curves of variable V18.
Note that data for $B$ and $I$ filters are shifted.}
\end{figure}

\clearpage

\begin{figure}
\figurenum{11}
\plotone{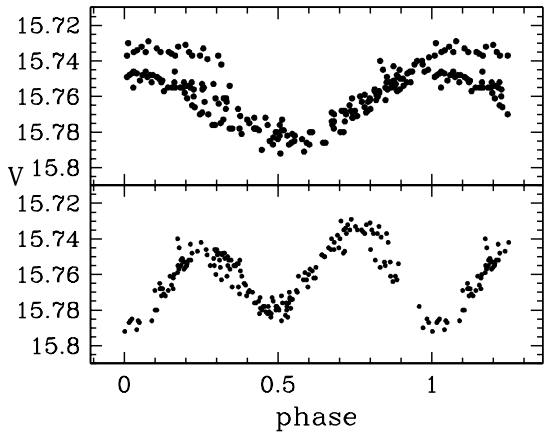}
\caption{The $V$-band light curve of V20 phased with a period
of 0.860 days (bottom) and with a period of 0.436 days.}
\end{figure}

\clearpage

\begin{figure}
\figurenum{12}
\plotone{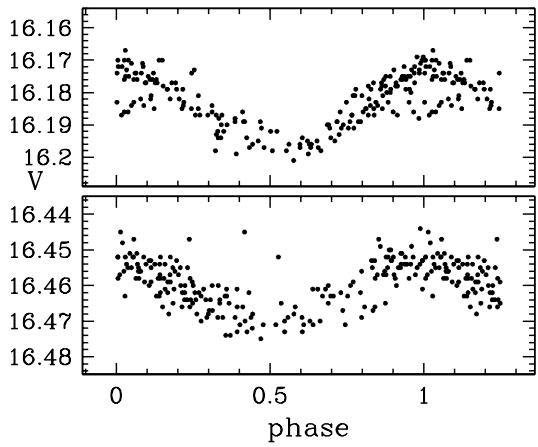}
\caption{Phased $V$ light curves of  V17 (upper) and V24.}
\end{figure}

\clearpage
\begin{figure}
\figurenum{13}
\plotone{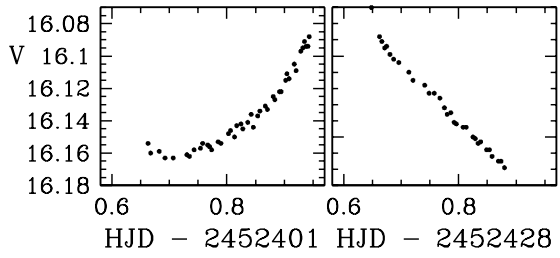}
\caption{Light curve of V22 for the nights of UT May 7 and UT June 2, 2002}
\end{figure}

\clearpage

\renewcommand{\arraystretch}{.6}

\begin{table}[ht]
\caption[]{\sc VARIABLE STARS IN THE CENTRAL REGION OF NGC~6397\\}
\begin{flushleft}
\begin{tabular}{lcccc}
\tableline\tableline
Name & R.A.      &Decl.        & HST        & ID \\
(1)  & (2)  & (3)       &(4)          &(5)            \\
\tableline
V7  & 17 40 43.74& -53 40 35.6 & W4(100,222)  & WF4-2\\
V10 & 17 40 37.43& -53 40 36.4 &              & BS11 \\
V11 & 17 40 43.95& -53 40 40.9 & W4(151,246)  & BS9\\
V12 & 17 40 41.42& -53 40 19.6 & W1(489,504)  & U23, CV1 \\
V13 & 17 40 48.82& -53 39 49.0 & W3(511,101)  & U10, CV6\\
V14 & 17 40 46.31& -53 41 15.9 & W4(548,348)  &   \\
V15 & 17 40 45.24& -53 40 25.2 & W4(122,53)   & BS10\\
V16 & 17 40 44.44& -53 40 42.0 & W4(190,223)  & MSP, WF4-1\\
V17 & 17 40 43.63& -53 41 16.8 & W4(384,522)  &  \\
V18 & 17 40 43.45& -53 40 28.1 & PC(345,91)   &   \\
V19 & 17 40 44.66& -53 40 23.8 & PC(386,266)  & PC-1\\
V20 & 17 40 41.51& -53 40 33.7 & PC(697,276)  & BS6\\
V21 & 17 40 41.40& -53 40 23.9 & PC(559,441)  & BS7\\
V22 & 17 40 41.02& -53 40 42.2 &              & BS8\\
V23 & 17 40 39.21& -53 40 46.9 &              & BS16\\
V24 & 17 40 38.97& -53 40 23.3 &              &  \\
\tableline\\
\end{tabular}
~\\
Note: Cols. (2)-(3): Units of right ascension are hours, minutes and seconds,
and units of declination are degrees, arcminutes, and arcseconds. Col. (5)
Pixel coordinates on the HST archive image u5dr0401r proceeded by
a name  of WFPC-2 camera CCD.  Col. (6) Other names of variables 
used in Grindlay et al. (2001), Taylor et al. (2001), 
and Lauzeral et al. (1992)
\end{flushleft}
\label{tab:phot1}
\end{table}

\clearpage

\begin{table}[ht]
\caption[]{\sc PHOTOMETRIC DATA FOR NGC 6397 VARIABLES\\}
\begin{flushleft}
\begin{tabular}{lcccccc}
\tableline\tableline
Name & Type & Period & $B$ & $V$ & $I$ & $\Delta V$ \\
\tableline
V7  & W~UMa  & 0.2699(2) & 17.72 & 17.05 & 16.11 & 0.47 \\
V10 & SX~Phe & 0.03006   & 16.36 & 15.97 & 15.46 & 0.12 \\
V11 & SX~Phe & 0.03826   & 15.78 & 15.40 & 14.88 & 0.05 \\
V12 & CV & 0.472(2)  & 18.51 & 17.95 & 16.96 & 0.37 \\
V13 & CV & 0.2352(5) & 20.04 & 19.35 & 18.27 & 0.45 \\
V14 & Eclipsing    & 0.3348(7) & 20.21 & 19.19 & 17.93 & 1.02 \\
V15 & SX~Phe & 0.02145   & 15.80 & 15.44 & 14.94 & 0.05 \\
V16 & Ell    & 1.35406   & 17.36 & 16.65 & 15.71 & 0.15 \\
V17 & $\gamma$~Dor? & 0.525(5)  & 16.81 & 16.17 & 15.32 & 0.025 \\
V18 & Eclipsing    & 0.7871(3) & 16.23 & 15.71 & 15.01 & 0.14 \\
V19 & W~UMA  & 0.2538(2) & 17.75 & 17.09 & 16.16 & 0.06 \\
V20 & W~UMa?, $\gamma$~Dor?  & 0.861(3)  & 16.22 & 15.75 & 15.08 & 0.04 \\
V21 & SX~Phe & 0.03896   & 15.88 & 15.48 & 14.91 & 0.30 \\
V22 & ?      & ?         & 16.61 & 16.16 & 15.58 & 0.11 \\
V23 & SX~Phe & 0.03717   & 16.05 & 15.66 & 15.14 & 0.04 \\
V24 & $\gamma$~Dor? & 0.457(2)  & 17.07 & 16.45 & 15.60 & 0.02 \\
\tableline
\end{tabular}
~\\
Note: Periods are given in days. $BVI$ magnitudes are given at maximum 
brightness with exception of SX~Phe stars for which average magnitudes
are listed. The last column gives the difference between observed extremes
for the $V$ light curves: $\Delta V = V_{\rm min}-V_{\rm max}$.
For CVs the quoted magnitudes refer to 8 nights from the beginning of UT May, 2002. 
\end{flushleft}
\label{tab:phot2}
\end{table}
\end{document}